\documentclass[5p]{elsarticle}

\usepackage{hyperref}
\usepackage{graphicx,color}
\usepackage{ulem}
\usepackage[T1]{fontenc}

\journal{Physics Letter B}

\begin{document}

\begin{frontmatter}

\title{Intruder configurations in $^{29}$Ne at the transition into the island of inversion: \\
Detailed structure study of $^{28}$Ne}


\author[riken_henp]{H. Wang\corref{mycorrespondingauthor}}
\cortext[mycorrespondingauthor]{Corresponding author}
\ead{he.wang@riken.jp}

\author[TiTech]{M. Yasuda}
\author[TiTech]{Y. Kondo}
\author[TiTech]{T. Nakamura}
\author[TiTech,Surrey]{J.\,A. Tostevin}
\author[Kyushu2,RCNP]{K. Ogata}
\author[Tokyo U,Nishina]{T. Otsuka}
\author[CSIC]{A. Poves}
\author[Tsukuba]{N. Shimizu}
\author[JAEA]{K. Yoshida}
\author[LPC]{N.L. Achouri}
\author[Lebanese]{H. Al Falou}
\author[TUD]{L. Atar}
\author[TUD,GSI]{T. Aumann}
\author[Nishina]{H. Baba}
\author[GSI]{K. Boretzky}
\author[GSI]{C. Caesar}
\author[Irfu]{D. Calvet}
\author[IBS]{H. Chae}
\author[Nishina]{N. Chiga}

\author[Irfu]{A. Corsi}
\author[LBNL]{H.L. Crawford}
\author[LPC]{F. Delaunay}

\author[Irfu]{A. Delbart}

\author[LPC]{Q. Deshayes}

\author[Atomki]{Zs. Dombr\'adi}

\author[KVI]{C. Douma}

\author[Atomki]{Z. Elekes}

\author[LBNL]{P. Fallon}

\author[Rud,GSI]{I. Ga\ifmmode \check{s}\else \v{s}\fi{}pari\ifmmode \acute{c}\else \'{c}\fi{}}

\author[Irfu]{J.-M. Gheller}

\author[LPC]{J. Gibelin}
\author[Irfu]{A. Gillibert}
\author[KVI]{M.N. Harakeh}
\author[TiTech]{A. Hirayama}
\author[Argonne]{C.R. Hoffman}
\author[TUD]{M. Holl}
\author[TUD]{A. Horvat}
\author[Eotvos]{\'A. Horv\'ath}
\author[IBS]{J.W. Hwang}
\author[Nishina]{T. Isobe}
\author[TUD]{J. Kahlbow}
\author[KVI]{N. Kalantar-Nayestanaki}
\author[Kyushu]{S. Kawase}
\author[IBS]{S. Kim}
\author[Nishina]{K. Kisamori}
\author[Tohoku]{T. Kobayashi}
\author[GSI]{D. K\"orper}
\author[Tokyo U]{S. Koyama}
\author[Atomki]{I. Kuti}
\author[Irfu]{V. Lapoux}
\author[Chalmers]{S. Lindberg}
\author[LPC]{F.M. Marqu\'es}
\author[CNS]{S. Masuoka}
\author[Koln]{J. Mayer}
\author[NSCL]{K. Miki}
\author[Kyoto U]{T. Murakami}
\author[KVI]{M.A. Najafi}
\author[Kyushu]{K. Nakano}
\author[Kyoto U]{N. Nakatsuka}
\author[Chalmers]{T. Nilsson}
\author[Irfu]{A. Obertelli}
\author[LPC]{N.A. Orr}
\author[Nishina]{H. Otsu}
\author[TiTech]{T. Ozaki}
\author[Nishina]{V. Panin}
\author[TUD]{S. Paschalis}
\author[GANIL,LPC,Irfu]{A. Revel}
\author[TUD,GSI]{D. Rossi}
\author[TiTech]{A.T. Saito}
\author[Tokyo U]{T. Saito}
\author[Nishina]{M. Sasano}
\author[Nishina]{H. Sato}
\author[SNU]{Y. Satou}
\author[TUD]{H. Scheit}
\author[TUD]{F. Schindler}
\author[CNS]{P. Schrock}
\author[TiTech]{M. Shikata}
\author[Nishina]{Y. Shimizu}
\author[GSI]{H. Simon}
\author[Atomki]{D. Sohler}
\author[GANIL]{O. Sorlin}
\author[IBS,Nishina]{L. Stuhl}
\author[TiTech]{S. Takeuchi}
\author[Osaka]{M. Tanaka}
\author[NSCL,NSCL2]{M. Thoennessen}
\author[TUD,GSI]{H. T\"ornqvist}
\author[TiTech]{Y. Togano}
\author[TiTech]{T. Tomai}
\author[TUD]{J. Tscheuschner}
\author[TiTech]{J. Tsubota}
\author[Nishina]{T. Uesaka}
\author[Nishina]{Z. Yang}
\author[Nishina]{K. Yoneda}

\address[riken_henp]{RIKEN, High Energy Nuclear Physics Laboratory, 2-1 Hirosawa, Wako, Saitama 351-0198, Japan}
\address[TiTech]{Department of Physics, Tokyo Institute of Technology, 2-12-1 O-Okayama, Meguro, Tokyo 152-8551, Japan}
\address[Surrey]{Department of Physics, Faculty of Engineering and Physical Sciences, University of Surrey, Guildford, Surrey GU2 7XH, United Kingdom}
\address[Kyushu2]{Department of Physics, Kyushu University, Fukuoka 819-0395, Japan}
\address[RCNP]{Research Center for Nuclear Physics (RCNP), Osaka University, Ibaraki 567-0047, Japan}
\address[Tokyo U]{University of Tokyo, Tokyo 1130033, Japan}
\address[Nishina]{RIKEN Nishina Center, Hirosawa 2-1, Wako, Saitama 351-0198, Japan}
\address[CSIC]{Departamento de F\'isica Te\'orica and IFT UAM-CSIC, Universidad Aut\'onoma de Madrid, 28049 Madrid, Spain}
\address[Tsukuba]{Center for Computational Sciences, University of Tsukuba, 1-1-1 Tennodai, Tsukuba 305-8577, Japan}
\address[JAEA]{Advanced Science Research Center, Japan Atomic Energy Agency, Tokai, Ibaraki 319-1195, Japan}
\address[LPC]{LPC Caen, CNRS/IN2P3, ENSICAEN, UniCaen, Normandie Universit\'e, 14050 CAEN Cedex, France}
\address[Lebanese]{Lebanese University, Beirut, Lebanon}
\address[TUD]{Institut f\"ur Kernphysik, Technische Universit\"at Darmstadt, 64289 Darmstadt, Germany}
\address[GSI]{GSI Helmholtzzentrum f\"ur Schwerionenforschung, 64291 Darmstadt, Germany}
\address[Irfu]{Irfu, CEA, Universit\'e Paris-Saclay, 91191 Gif-sur-Yvette, France}
\address[IBS]{Center for Exotic Nuclear Studies, Institute for Basic Science, Daejeon 34126, Republic of Korea}
\address[LBNL]{Nuclear Science Division, Lawrence Berkeley National Laboratory, Berkeley, California 94720, USA}
\address[Atomki]{Institute of Nuclear Research, Atomki, 4001 Debrecen, Hungary}
\address[KVI]{ESRIG, University of Groningen, Zernikelaan 25, 9747 AA Groningen, The Netherlands}
\address[Rud]{Ru{\dj}er Bo\ifmmode \check{s}\else \v{s}\fi{}kovi\'c Institute, HR-10002 Zagreb, Croatia}
\address[Argonne]{Physics Division, Argonne National Laboratory, Argonne, Illinois 60439, USA}
\address[Eotvos]{E\"otv\"os Lor\'and University, P\'azm\'any P\'eter S\'et\'any 1/A, H-1117 Budapest, Hungary}
\address[Kyushu]{Department of Advanced Energy Engineering Science, Kyushu University, Kasuga, Fukuoka 816-8580, Japan}
\address[Tohoku]{Department of Physics, Tohoku University, Miyagi 980-8578, Japan}
\address[Chalmers]{Institutionen f\"or Fysik, Chalmers Tekniska H\"ogskola, 412 96 G\"oteborg, Sweden}
\address[CNS]{Center for Nuclear Study, University of Tokyo, 2-1 Hirosawa, Wako, Saitama 351-0198, Japan}
\address[Koln]{Institut f\"ur Kernphysik, Universit\"at zu K\"oln, 50937 K\"oln, Germany}
\address[NSCL]{Facility for Rare Isotope Beams, Michigan State University, East Lansing, Michigan 48824, USA}
\address[Kyoto U]{Department of Physics, Kyoto University, Kyoto 606-8502, Japan}
\address[GANIL]{Grand Acc\'el\'erateur National d’Ions Lourds (GANIL), CEA/DRF-CNRS/IN2P3, Bvd Henri Becquerel, 14076 Caen, France}
\address[SNU]{Department of Physics and Astronomy, Seoul National University, 1 Gwanak-ro, Gwanak-gu, Seoul 08826, Korea}
\address[Osaka]{Department of Physics, Osaka University, Osaka 560-0043, Japan}
\address[NSCL2]{Department of Physics and Astronomy, Michigan State University, East Lansing, Michigan 48824, USA}

\begin{abstract}
Detailed $\gamma$-ray spectroscopy of the exotic neon isotope $^{28}$Ne has been performed for the first time 
using the one-neutron removal reaction from $^{29}$Ne on a liquid hydrogen target at 240~MeV/nucleon.
Based on an analysis of parallel momentum distributions, 
a level scheme with spin-parity assignments has been constructed for $^{28}$Ne and the negative-parity states are identified for the first time. 
The measured partial cross sections and momentum distributions reveal a significant intruder $p$-wave strength providing evidence of the breakdown of the $N=20$ and $N=28$ shell gaps.
Only a weak, possible $f$-wave strength was observed to bound final states. 
Large-scale shell-model calculations with different effective interactions do not reproduce the large $p$-wave and small $f$-wave strength observed experimentally,
indicating an ongoing challenge for a complete theoretical description of the transition into the island of inversion along the Ne isotopic chain.
\end{abstract}


\end{frontmatter}



The understanding of nuclei far from the line of $\beta$ stability is one of the key challenges of modern nuclear physics. 
It is predicted that these exotic nuclei, with large imbalances between proton ($Z$) and neutron ($N$) numbers, 
will undergo dramatic changes to their shell structures compared to those established for stable nuclei. 
Indeed, such modifications of shell structure have now been observed for a wide range of exotic nuclei, 
revealing the breakdown of conventional shell gaps and the formation of new ones.

A pivotal area for a sudden shell structure change is the so-called "island of inversion", 
comprising neutron-rich F($Z=9$), Ne ($Z=10$), Na ($Z=11$), and Mg ($Z=12$) isotopes around $N=20$. 
The first such indication, from mass measurements of $^{31,32}$Na~\cite{Thibault_31Na}, 
found these systems to be more bound than expected for a spherical shape with the conventional $N=20$ shell gap. 
An onset of quadrupole deformation was later reported in $^{32}$Mg~\cite{Motobayashi_32Mg} and $^{30}$Ne~\cite{Yanagisawa_30Ne, Doornenbal_30Ne},
revealed by the low excitation energies of their first $2^+_1$ states and large $B(E2;0_{gs} \rightarrow 2^+_1)$ transition probabilities. 
Moving towards the more neutron-rich isotopes, large quadrupole deformation was found in $^{40}$Mg at $N=28$~\cite{Crawford_40Mg}. 
As modeled by shell-model calculations~\cite{Otsuka_rev}, this dramatic change in nuclear structure in the island of inversion 
is attributed to neutron particle-hole ($n$p-$n$h) excitations across the quenched $N=20$ and $N=28$ shell gaps between the $sd$ and $pf$ orbitals.

Many experimental efforts have been made to explore the mechanism driving the emergence of the island of inversion. 
While extensive studies have been performed for the magnesium isotopic chain~\cite{Terry_30Mg, Fer30Mg, Kitamura_30Mg, Bazin_33Mg, Kitamura_32Mg,Kitamura_32Mg_2}, 
such detailed structure information remains more limited for the Ne isotopes. 
Along the Ne isotopic chain, the first onset of a $pf$-shell intruder configuration was identified in $^{28}$Ne~\cite{Terry_28Ne}, 
while $^{29}$Ne~\cite{Liu_29Ne, Kob29, Holl_29Ne,Revel_29Ne} borders, and $^{30-32}$Ne~\cite{Fallon_30Ne,Doornenbal_32Ne,Nakamura_31Ne,Murray_32Ne} lie within the island of inversion.
Experimental findings suggest that the intruder $p$-wave configuration plays a key role.
The spectroscopic factor (C$^2$S) for removing a neutron from the 2$p_{3/2}$ intruder state in $^{28}$Ne was deduced to be 0.32(4)~\cite{Terry_28Ne}. 
This 2$p_{3/2}$ neutron removal strength increases to 0.9(1) in $^{30}$Ne~\cite{Liu_29Ne}. 

For $^{29}$Ne, a dominant $p$-wave ground-state configuration was deduced by measuring both
the inclusive and Coulomb dissociation cross sections for $^{28}$Ne production via the neutron removing reactions on a carbon and lead target, respectively~\cite{Kob29}.
In addition to these $p$-wave contributions, $f$-wave intruder neutron-removal strength
has been identified, populating unbound states in $^{27}$Ne~\cite{Brown_27Ne} and $^{29}$Ne~\cite{Holl_29Ne}. 
These $f$-wave components reveal a markedly different picture, compared to shell-model calculations, 
which predict significant $f$-wave intruder strength leading to bound final states~\cite{Liu_29Ne, Brown_27Ne}. 
Obtaining a quantitative determination of this $pf$-shell strength in the ground state of $^{29}$Ne, 
at the transition into the island of inversion, is critical to understand the development of such intruder configurations along the neon isotopic chain -- as well as to
provide essential input for developing its theoretical description.

Accordingly, the neutron-rich Ne isotopes can provide a testing ground offering new insight into this theoretical description. 
Originally, models suggested that the structures developing in the island-of-inversion nuclei were expected to be dominated by 2p-2h intruder configurations~\cite{SDPF-M}. 
However, more recent theoretical studies have proposed considerable 4p-4h intruder contributions~\cite{SDPF-U-MIX,EEdf1,three_level} 
to explain the evolution of intruder $pf$ strength from $^{30}$Mg~\cite{Fer30Mg} to $^{32}$Mg~\cite{three_level,Elder_32Mg}. 
For the neon isotopes, the 4p-4h configuration was hinted by the suppression of the two-proton removal cross section from $^{30}$Ne~\cite{Fallon_30Ne} 
whereas a very recent study of $^{32}$Ne found no significant difference in expectations between theoretical descriptions that include only 2p-2h and those with 4p-4h configurations~\cite{Murray_32Ne}. 
More detailed spectroscopic information in the Ne isotopes is needed to pin down these different descriptions of the island of inversion.

In this work, we report on in-beam $\gamma$-ray spectroscopy of $^{28}$Ne ($Z=10,N=18$) produced by the one-neutron removal reaction from $^{29}$Ne on a proton target. 
These new experimental results provide a more detailed level scheme, 
via the partial cross sections to the individual bound states in $^{28}$Ne and their momentum distributions -- a direct probe of the active, valence neutron orbitals. 
This allows a quantitative determination of the strength of $pf$ intruder configurations in the $^{29}$Ne ground state for the first time.

A hydrogen target was used due to the advantages of proton-induced reactions 
in nuclear structure study with exotic beams~\cite{proton_target}, such as high luminosity.
Detailed comparisons of analyses of specific nucleon removal experiments on proton and light nuclear targets have not yet been made. 
One purpose of the present hydrogen target experiment is to provide such data 
to compare with the earlier comprehensive (but less exclusive) carbon target data~\cite{Kob29} and analysis 
and to begin a program of such comparisons.

Located at the border of the island of inversion, $^{28}$Ne is expected to demonstrate intruder $pf$ strength. 
To date, studies of $^{28}$Ne have mainly concerned the properties of the $2^+_1$ state~\cite{Pritychnko_28Ne,Iwasaki_28Ne,Dombradi_28Ne,Michimasa_28Ne}. 
Along the Ne isotopic chain, a lowering of the $2^+_1$ state energy, $E_x(2^+_1)$, occurs at $^{28}$Ne, suggesting an onset of quadruple deformation~\cite{Pritychnko_28Ne}. 
However, the deformation parameters, for both protons and neutrons~\cite{Iwasaki_28Ne,Dombradi_28Ne,Michimasa_28Ne} were found to be moderate as compared to other nuclei inside the island of inversion. 
On the other hand, the available spectroscopic studies on the $N=18$ isotones have revealed the development of intruder $pf$ configurations --
with the intruder contributions present in low-lying excited states and expected to be present even in ground states towards the low-$Z$ part of the chain -- due to shell evolution. 
Indeed, shell-model calculations also suggest that the $N=20$ and $N=28$ shell gaps will reduce with decreasing atomic number~\cite{Utsuno_N28gap}. 
For instance, intruder states were found in the excited $0^+$ states in $^{34}$Si ($Z=14$)~\cite{Rotaru_34Si} and $^{30}$Mg ($Z=12$)~\cite{Schwerdtfeger_30Mg}
as well as in the low-lying excited states in $^{29}$Na ($Z=11$)~\cite{Tripathi_29Na}. 
Moving to $^{28}$Ne, additional contributions from intruder configurations are expected according to shell-model calculations~\cite{SDPF-M,EEdf1,SDPF-U-MIX} as compared to the heavier $N=18$ isotones.
Thus, detailed structural information will allow quantitative study of this $pf$-shell intruder strength in $^{28}$Ne.

\begin{figure}[th]
\begin{center}
  \includegraphics[width=12cm]{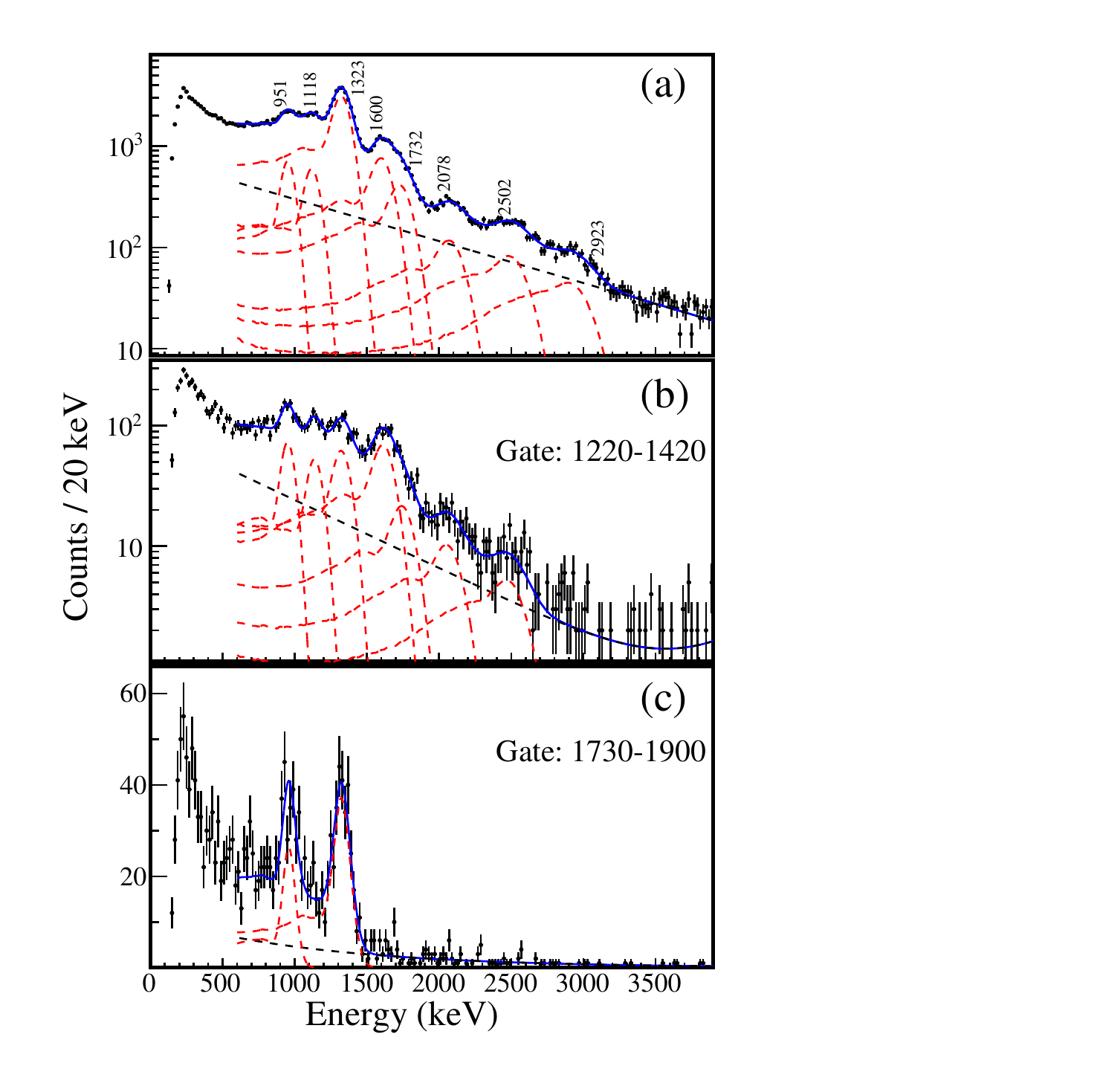}
\caption{\label{fig:spectrum}
 (a) $\gamma$-ray energy spectrum in coincidence with the $^{29}$Ne$(p,pn)$$^{28}$Ne reaction, after Doppler-shift correction.
The fit result (blue line) includes simulated DALI2 response functions (red dashed lines) and a double-exponential background (black dashed line).
The $\gamma$-ray spectra are shown for a gate on the 1323-keV (b) and 1732-keV (c) transitions. 
The self-coincidences in (b) originate from the Compton events of transitions at high energy. 
The gate used for $\gamma-\gamma$ analysis in (c) was shifted from the centroid of the 1732-keV peak to minimize the cross contamination from overlapping peaks. 
 }
\end{center}
\end{figure}

\begin{figure}[th]
\begin{center}
  \includegraphics[width=8.9cm]{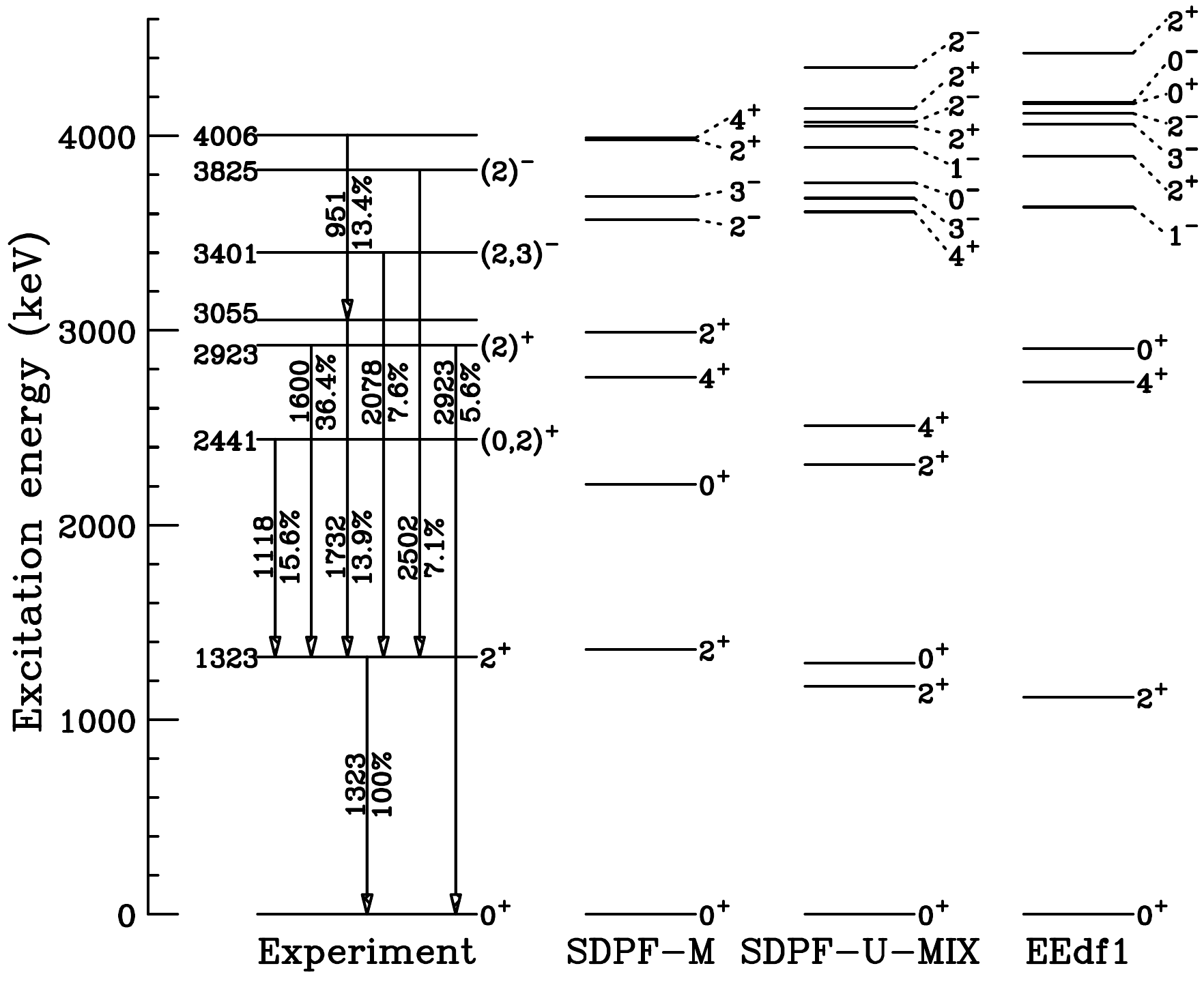}
\caption{\label{fig:ne28_level}
Experimental and calculated level schemes for $^{28}$Ne. 
Excitation energies, spins and parities are provided beside the levels. 
Observed $\gamma$-ray energies and their relative intensities are also given.
The experimental spin-parity assignments are based on the analysis on the shapes of the parallel momentum distribution obtained in coincidence with the transitions. 
Three sets of the calculated level schemes using different shell-model interactions SDPF-M, SDPF-U-MIX and EEdf1 are shown for comparison. See details in the text.
 }
\end{center}
\end{figure}

The experiment was performed at the Radioactive Isotope Beam Factory (RIBF),
operated by the RIKEN Nishina Center and the Center for Nuclear Study, University of Tokyo. 
A $^{29}$Ne secondary beam was produced by fragmentation of a $^{48}$Ca primary beam at 345~MeV/nucleon, with a typical intensity of 550~pnA, 
on a 15-mm-thick beryllium target located at the entrance of the BigRIPS fragment separator~\cite{Kubo_BigRIPS_PTEP}.
The $^{29}$Ne beam was selected and purified in BigRIPS by employing two wedge-shaped aluminum degraders at the dispersive foci. 
Identification of the $^{29}$Ne particles was made event-by-event by measuring the time of flight (TOF), magnetic rigidity ($B\rho$), and energy loss ($\Delta E$), as described in Ref.~\cite{Fukuda_PID}. 
The average intensity of the $^{29}$Ne beam in BigRIPS was 8.2 $\times$ 10$^3$ particles per second and its fraction in the secondary beams was $\sim$85\%. 

The $^{29}$Ne beam impinged on the MINOS~\cite{MINOS} target, a 151(1)-mm-thick liquid hydrogen target, 
with an energy of 240~MeV/nucleon at the center of the target.
The target was surrounded by a time projection chamber for measuring the trajectory of the recoil proton from the ($p,pn$) reaction.
The tracks of the beam particles, measured by the drift chamber before the target, and the proton were used to reconstruct the reaction vertex in the target~\cite{MINOS}. 
The measured efficiency to detect the outgoing proton was 67(5)\%.

To detect the $\gamma$ rays emitted from the excited states of $^{28}$Ne, 
the MINOS target was surrounded by the DALI2 array~\cite{Takeuchi_DALI2}, which consisted of 142~NaI(Tl) scintillators.
The efficiency and energy resolution for 1-MeV $\gamma$ rays were 15\% and 11\% (FWHM), respectively.
The DALI2 response to $\gamma$ rays was generated using Monte Carlo simulations with the GEANT4~\cite{GEANT4} framework and the experimental $\gamma$-ray energy spectrum was fitted with these response functions.
DALI2 was chosen not only for its high $\gamma$-ray detection efficiency
but also because the MINOS setup is optimized for coupling with it in a compact geometry~\cite{MINOS}.

Downstream of MINOS, the reaction residues were transported to the SAMURAI spectrometer~\cite{SAMURAI}
and identified by measuring their $B\rho$, $\Delta E$ and TOF information. 
A central magnetic field of 2.9~Tesla was applied for the SAMURAI dipole magnet. 
The $B\rho$ values of charged particles were reconstructed using two drift chambers located at the entrance and exit of the magnet~\cite{SAMURAI}. 
The TOF and  $\Delta E$ measurements were made using a 24 element plastic scintillator hodoscope.

The Doppler-shift-corrected $\gamma$-ray energy spectrum measured in coincidence with the $^{29}$Ne$(p,pn)$$^{28}$Ne channel is shown in Fig.~\ref{fig:spectrum}(a).
The spectrum was fitted using simulated DALI2 response functions added to a double-exponential background. 
Eight transitions were identified, as listed in Table~\ref{table_1}.
The 1323-keV transition with the most intensity, corresponding to the $2^+_1 \rightarrow 0^+_{gs}$ decay, 
is consistent with previous reports~\cite{Kob29,Pritychnko_28Ne,Iwasaki_28Ne,Dombradi_28Ne, Michimasa_28Ne,Fallon_28Ne,Belleguic_28Ne}.
The second most intense transition is found at 1600~keV, confirming the previous report~\cite{Kob29}. 
The peak at 951~keV confirms the previous measurements~\cite{Kob29, Belleguic_28Ne}.
The transitions at 2078, 2502, and 2923 keV are newly found in the present work. 
$\gamma-\gamma$ analysis showed that the transitions at 951, 1118, 1600, 1732, 2078, and 2502~keV are in coincidence with the $2^+_1 \rightarrow 0^+_{gs}$ decay, as displayed in Fig.~\ref{fig:spectrum}(b).
Within the energy uncertainties, the 2923-keV transition is in good agreement with the sum of the 1323- and 1600-keV transitions. 
The 1118-keV transition, reported in Refs.~\cite{Fallon_28Ne, Rodr_28Ne} at 1127~keV, is suggested to be in coincidence with the $2^+_1 \rightarrow 0^+_{gs}$ decay, based on this neutron removal reaction data.
Gating around 1732~keV, peaks are found at 951~keV and 1323~keV, as shown in Fig.~\ref{fig:spectrum}(c), suggesting a $\gamma$-ray cascade.
The 951- and 1732-keV transitions are found to have the same intensities in the present work and thus their placement is difficult. 
The cascade decay may correspond to those reported in the inelastic scattering~\cite{Michimasa_28Ne} and one-proton knockout reactions~\cite{Fallon_28Ne},
where the 1732-keV transition is more intense than the 951-keV one. 
In addition, a large intensity for the 1732-keV transition is suggested from other measurements~\cite{Dombradi_28Ne, Belleguic_28Ne}.
Therefore, the 1732-keV transition is tentatively placed lower than the 951-keV transition. 
Based on the observed $\gamma-\gamma$ coincidences, a level scheme is constructed as presented in Fig.~\ref{fig:ne28_level}.

The measured inclusive cross section to all bound states was determined to be 28(1)~mb.
As a thick target is used, the reaction loss in the thick liquid hydrogen target is accounted for using the method presented in Ref.~\cite{Kob29}. 

Using the fit result and level scheme constructed above, the partial cross sections associated with each excited state were obtained. 
These partial cross sections are summarized in Table~\ref{table_2}, ordered by increasing excitation energy.
The ground-state cross section was obtained by subtracting the cross section to all excited states from the inclusive one.

\begin{table}[h]
\caption{Observed $\gamma$-ray energies ($E_{\gamma}$) from the one-neutron knockout reaction from $^{29}$Ne and their placements in $^{28}$Ne.
Intensities relative to the 1323-keV transition are also shown. 
The uncertainties in the relative intensities indicate the statistical contributions.
\label{table_1}
}
\centering
\begin{tabular}{lcc}
\\
\hline
$E_{\gamma}$ (keV) & Placement & Relative intensity (\%) \\
\hline
951(5)      & 4006 $\rightarrow$ 3055 & 13.4(1)  \\
1118(6)    & 2441 $\rightarrow$ 1323 & 15.6(1)   \\
1323(7)    & 1323 $\rightarrow$ 0 & 100.0(2)   \\
1600(9)    & 2923 $\rightarrow$ 1323 & 36.4(2)  \\
1732(10)  & 3055 $\rightarrow$ 1323 & 13.9(1)  \\
2078(12)  & 3401 $\rightarrow$ 1323 & 7.6(4)   \\
2502(17)  & 3825 $\rightarrow$ 1323 & 7.1(5)    \\
2923(22)  & 2923 $\rightarrow$  0 & 5.3(4)    \\
\hline
\end{tabular}
\end{table}

\begin{table*}[h]
\caption{Results for one-neutron removal from $^{29}$Ne. 
Level energies ($E_{x}$) and partial cross sections ($\sigma_{\rm exp}$) are listed.  
The orbitals of the removed neutron are also shown. 
Calculated theoretical single-particle cross sections ($\sigma_{\rm sp}$) using the DWIA~\cite{DWIA_1,DWIA_2} and eikonal frameworks~\cite{eikonal_arxiv} are presented in the last two columns. 
The experimental spectroscopic factors can be obtained by using C$^2$S$_{\rm exp-DWIA(eikonal)}$ = $\sigma_{\rm exp}$/$\sigma_{\rm sp-DWIA(eikonal)}$.
\label{table_2}
}
\centering
\begin{tabular}{lcccccc}
\\
\hline
$E_x$ (keV) & $\sigma_{\rm exp}$ (mb) & orbital & $\sigma_{\rm sp-DWIA}$ (mb) & $\sigma_{\rm sp-eikonal}$ (mb) \\
\hline
     0 &   8.4(8) & $2p_{3/2}$  & 14.81 &  21.19 \\
1323 &  3.6(3) & $2p_{3/2}$$^a$  & 11.93  & 18.56  \\
         &            & $2p_{3/2}$$^b$  & 11.93  & 18.56  \\
         &            & $1f_{7/2}$$^b$   &   8.00  &  13.69  \\
2441 &  2.9(1) & $2p_{3/2}$  & 10.57 &  17.18  \\
2923 &  7.7(2)  & $2p_{3/2}$  & 10.13 & 16.70  \\
3055 &  0.1(1) &                    &             &             \\
3401 &  1.4(1) & $1d_{3/2}$$^c$  & 5.90     & 11.46   \\
         &             & $2s_{1/2}$$^d$  & 8.46    &    \\
         &            & $1p_{3/2}$$^e$  & 9.74    &    \\
3825 &  1.3(1) & $2s_{1/2}$  & 8.14     &  14.80   \\
4006 &  2.5(1) & $1d_{3/2}$  & 5.70     &  11.13    \\
Inclusive &     28(1) &                    &             &  \\
\hline
\end{tabular}\\
\footnotesize{$^a$ A removal from $2p_{3/2}$ is assumed.}  \\
\footnotesize{$^b$ A linear combination of $2p_{3/2}$ and $1f_{7/2}$ is assumed. }  \\
\footnotesize{$^c$ A removal from $1d_{3/2}$ is assumed.}  \\
\footnotesize{$^d$ A removal from $2s_{1/2}$ is assumed.}  \\
\footnotesize{$^e$ A removal from $2p_{3/2}$ is assumed.} 
\end{table*}

\begin{figure}
   \includegraphics[width=10.5cm]{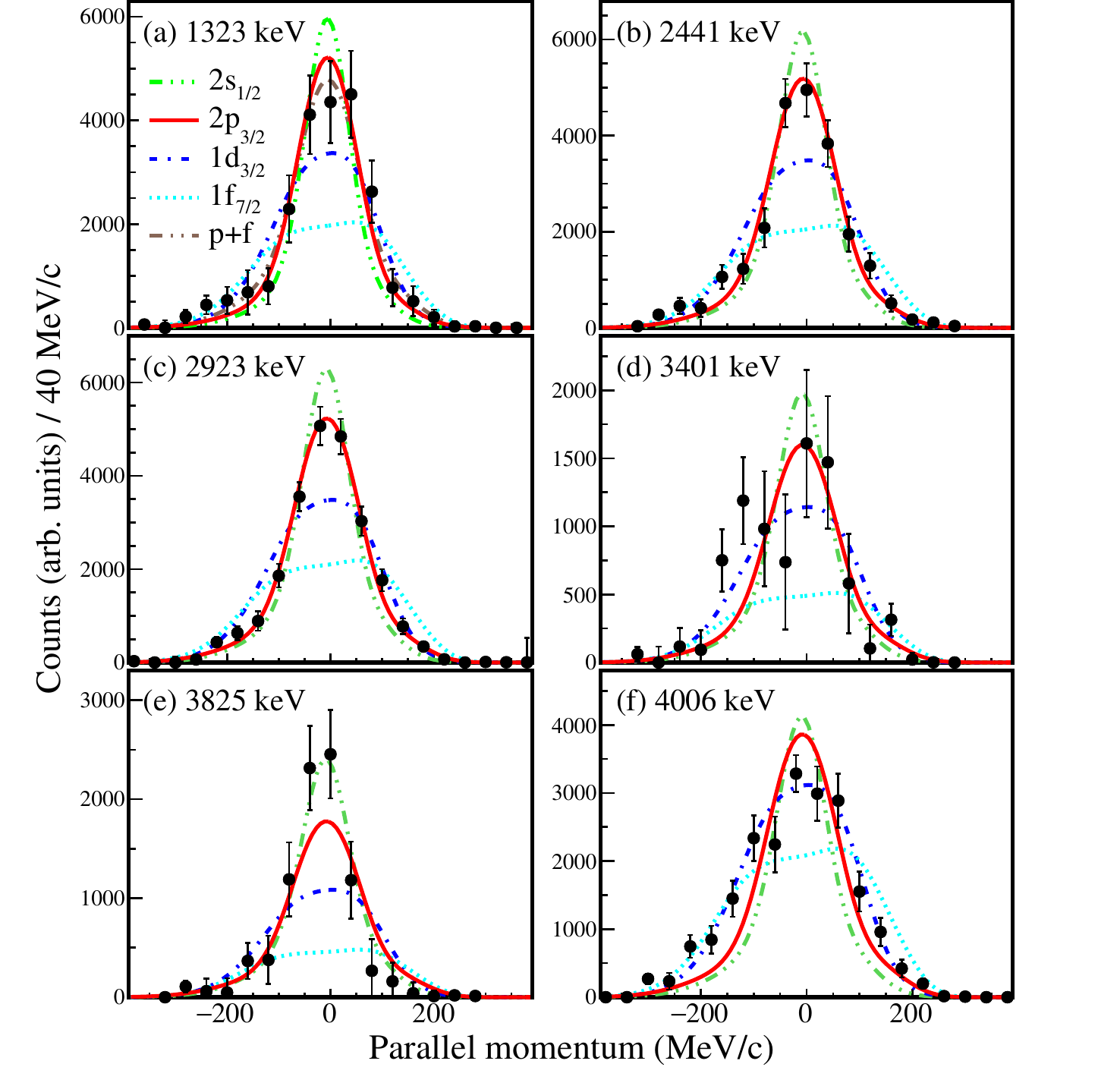}
\caption{\label{fig:momentum}
Measured parallel momentum distributions for the $^{28}$Ne residues obtained following the $^{29}$Ne($p,pn)$ in coincidence with $\gamma$-ray transitions. 
The distribution of the 1323-keV state was obtained by subtracting the contributions from the feeding states. 
The experimental data are compared with the shapes computed from the DWIA calculations
when assuming neutron-removal from the 2$s_{1/2}$ (green), 2$p_{3/2}$ (red) , 1$d_{3/2}$ (blue), and 1$f_{7/2}$ (cyan) orbitals. 
}
\end{figure}

The intermediate energy one-neutron removal reaction is known to be a powerful tool to assign spins and parities 
since the shapes of the residue momentum distributions reflect the orbital angular momentum of the removed neutron. 
To make the excited states spin-parity assignments, 
the shapes of the momentum distributions of the $^{28}$Ne residues in the $^{29}$Ne rest frame were analyzed. 
The parallel momentum distributions were reconstructed using the velocities of the beam and fragments, 
as well as the scattering angle determined from the reaction vertex and the drift chambers located behind the secondary target. 
A resolution of 30~MeV/$c$ (root mean square) was evaluated by measuring the unreacted $^{29}$Ne beam.
The momentum distributions for the excited states were determined by fitting the $\gamma$-ray spectra in coincidence with the selection of 40 MeV/c wide bins of the inclusive momentum
and obtaining the integral for each $\gamma$-ray transition. 

For the 1323-keV state, the momentum distribution was obtained after the subtraction of the contributions from the feeding states, based on the level scheme presented in Fig.~\ref{fig:ne28_level}. 
The momentum distribution for the 3055-keV state could not be obtained in the present work because its cross section is consistent with zero. 
The resulting momentum distributions are displayed in Fig.~\ref{fig:momentum}.
The large error bars in the 1323-keV momentum distribution are due to the subtraction of the feeding states.
For the 3401- and 3825-keV states, 
the errors in the momentum distributions are dominated by the statistical contributions because of their relatively small intensities.

These experimental momentum distributions are compared to those calculated using the distorted
wave impulse approximation (DWIA) model~\cite{DWIA_1,DWIA_2} assuming neutron removal from different
single-particle orbitals, namely 2$s_{1/2}$, 2$p_{3/2}$, 1$d_{3/2}$, and 1$f_{7/2}$. 
The model has been already applied in several works~\cite{Chen_53Ca,Taniuchi_78Ni, Olivier_79Cu, Yoshida_DWIA, Enciu_52Ca}.
For the comparison, the overall normalization factor in the calculated momentum distribution was treated as a free parameter.
In the DWIA approach, the single-particle wave function and nuclear density were derived from the Bohr-Mottelson single-particle potential~\cite{Bohr_Mottelson}.
The depth of the potential is adjusted to reproduce the neutron separation energy of $^{29}$Ne for each state of the $^{28}$Ne core.
Optical potentials used for the distorted waves in the initial and final states were constructed by the microscopic folding model~\cite{DWIA_folding} from the calculated nuclear density and the Melbourne $g$-matrix interaction~\cite{Melbourne}.
The Franey-Love parameterization~\cite{Franey_Love} was applied for the neutron-removing proton-neutron interaction of the elementary process. 
We note that the momentum distributions computed with the DWIA framework present small asymmetric shapes~\cite{DWIA_2}, and those from the eikonal dynamical model have symmetric ones~\cite{eikonal_arxiv}.
The widths of the distributions calculated from both theoretical approaches are essentially identical 
and our conclusions regarding the deduced spin-parities of final states are common to both methods.

The state at 1323~keV has been assigned to the $2^+_1$ state in the previous reports. 
Considering the ground state of $^{29}$Ne is $3/2^-$~\cite{Liu_29Ne, Kob29}, 
spin-parity conservation dictates that population of the $2^+_1$ state occurs due to neutron removal from the 2$p_{3/2}$ or 1$f_{7/2}$ orbitals. 
The measured 1323-keV momentum distribution is thus fitted assuming two possibilities: 
(i) removal from pure 2$p_{3/2}$, and (ii) removal from a linear combination of 2$p_{3/2}$ and 1$f_{7/2}$ orbitals, as shown in Fig.~\ref{fig:momentum}(a). 
The latter fit comprises $71 \pm 15 \%$ of $p$- and $29 \pm 15 \%$  of $f$-wave strength. 
Due to the size of the error bars after feeding subtraction, these two fits cannot be resolved. 
The fit using a combination of $p$- and $f$-wave configurations indicates the possibility of a small $f$-wave contribution to the bound states. 
Using this combination, the cross sections for the $p$- and $f$-wave strengths are 2.6(6) and 1.0(5)~mb.
For all other excited states we compare the momentum distributions to the shapes calculated for removal from a single orbital component.

As shown in Fig.~\ref{fig:momentum}(b), the momentum distribution for the 2441-keV state favors $p$-wave removal, for which spin-parities of $0^+,1^+,2^+,3^+$ are possible. 
A $3^+$ assignment is not likely, because the $3^+$ states normally appear at excitation energies above 5~MeV~\cite{NNDC} for the neutron-rich nuclei around $N=20$. 
Given that the $\gamma$-decay proceeds solely to the $2^+_1$ level, a spin-parity of $(0, 2)^+$ is tentatively assigned to the 2441-keV state 
because a $1^+$ assignment would favor an $M1$ decay to the ground state. 
It is noted that the 1118-keV transition is also reported in the one-proton removal from $^{29}$Na~\cite{Fallon_28Ne}. 
Since the $^{29}$Na ground state is $3/2^+$~\cite{NNDC} this supports the positive-parity ($J^{\pi}=1^+,2^+,3^+,4^+$) assignment for the 2441-keV state 
populated following proton removal from the 1$d_{5/2}$ orbital. 
Considering both the one-proton and one-neutron removal cases, the 2441-keV state favors a $2^+$ assignment.

The momentum distribution of the 2923-keV state clearly favors the knockout of a neutron from the 2$p_{3/2}$ orbital, as displayed in Fig.~\ref{fig:momentum}(c). 
Because the 2923-keV state has one branch that directly decays to the ground state, the $0^+$ assignment is excluded. 
A $3^+$ assignment is not favored, since it would result in an $M3$ transition for the $2923 \rightarrow 0^+_{gs}$ decay with a half-life of a few hundreds nanoseconds, 
according to the Weisskopf estimation. 
Such a half-life is too long to be observed with the experimental technique used in the present work.
Considering that the $2923 \rightarrow 1323$ decay is more intense than that to the ground state, a $2^+$ assignment is made for this state, 
because a $1^+$ assignment would result in a strong $M1$ decay for the $2923 \rightarrow 0^+_{gs}$ transition due to the large energy difference. 

Negative-parity states in $^{28}$Ne can be populated by removing a neutron from $sd$ orbitals.
For the state at 3401~keV, a removal from 1$d_{3/2}$ is most likely given its smallest reduced $\chi^2$ value of 1.07, as shown in Fig.~\ref{fig:momentum}(d). 
However, the limited statistics cannot exclude the possibilities of $s$- (reduced $\chi^2=1.66$) or $p$-wave (reduced $\chi^2=1.44$) configurations.
Removal from 1$d_{3/2}$ or 1$s_{1/2}$ leads to a negative-parity with $J^{\pi}=0^-, 1^-, 2^-, 3^-$ or $J^{\pi}=1^-, 2^-$, respectively. 
Given its sole decay branch to the $2^+_1$ state, the spin-parity of $1^-$ is not likely as this would favor the decay to the ground state via an $E1$ transition.
In the region of nuclei around $N=20$, the $0^-$ states are typically located at excitation energies around 10~MeV~\cite{NNDC}.
Therefore, a spin-parity of $(2,3)^-$ is possible for the 3401-keV state.
In the case of removal from 2$p_{3/2}$, a positive-parity with $J^{\pi}=0^+, 2^+$ is possible, as per the discussion for the 2441-keV state,
since the observed decay is solely to the $2^+_1$ state.
Therefore, the 3401-keV state may have a spin-parity of $(2,3)^-$ or $(0,2)^+$.
For the 3825-keV state, the momentum distribution supports removal from the 2$s_{1/2}$ orbital
and a $2^-$ assignment is proposed here because the decay proceeds only to the $2^+_1$ level.
The 4006-keV momentum distribution appears broad and favors a removal from 2$d_{3/2}$, indicating a negative-parity assignment with $J^{\pi}=0^--3^-$.

As mentioned above, the present direct neutron-removal reaction of $^{29}$Ne on a proton target has also been studied with the eikonal dynamical model~\cite{eikonal_arxiv}.
This eikonal approach, widely used for reactions on light-nucleus targets~\cite{Bazin_33Mg,Kitamura_32Mg,Kitamura_32Mg_2,Kob29}, 
was also recently applied to the proton-induced $^{30}$Ne$(p,pn)$~\cite{Holl_29Ne}, $^{29}$Ne$(p,2p)$ and $^{29}$F$(p,pn)$~\cite{Revel_28F} reactions.
In the case of a proton target, the nucleon removal is due to nucleon-nucleon collisions, described in the model by their elastic S-matrix. 
Unlike for a nuclear target, where the inelastic (target excitation) removal mechanism dominates, on a proton target removal is due to the elastic target collisions. 
As used here, the proton-$^{28}$Ne optical potential is deduced from the $^{28}$Ne density, obtained from a spherical Skyrme SkX Hartree-Fock (HF) calculation~\cite{Skyrme_HF}. 
The eikonal methodology used is outlined in Refs.~\cite{Holl_29Ne, Revel_28F} and detailed in Ref.~\cite{eikonal_arxiv}. 
The single-particle cross sections computed using the DWIA and eikonal models differ, as are listed in Table~\ref{table_2}.
The difference might be related to the different inputs used in the DWIA and eikonal calculations for the $^{28}$Ne density and geometry of the neutron bound state potentials as described above.

\begin{figure}
  \includegraphics[width=8.6cm]{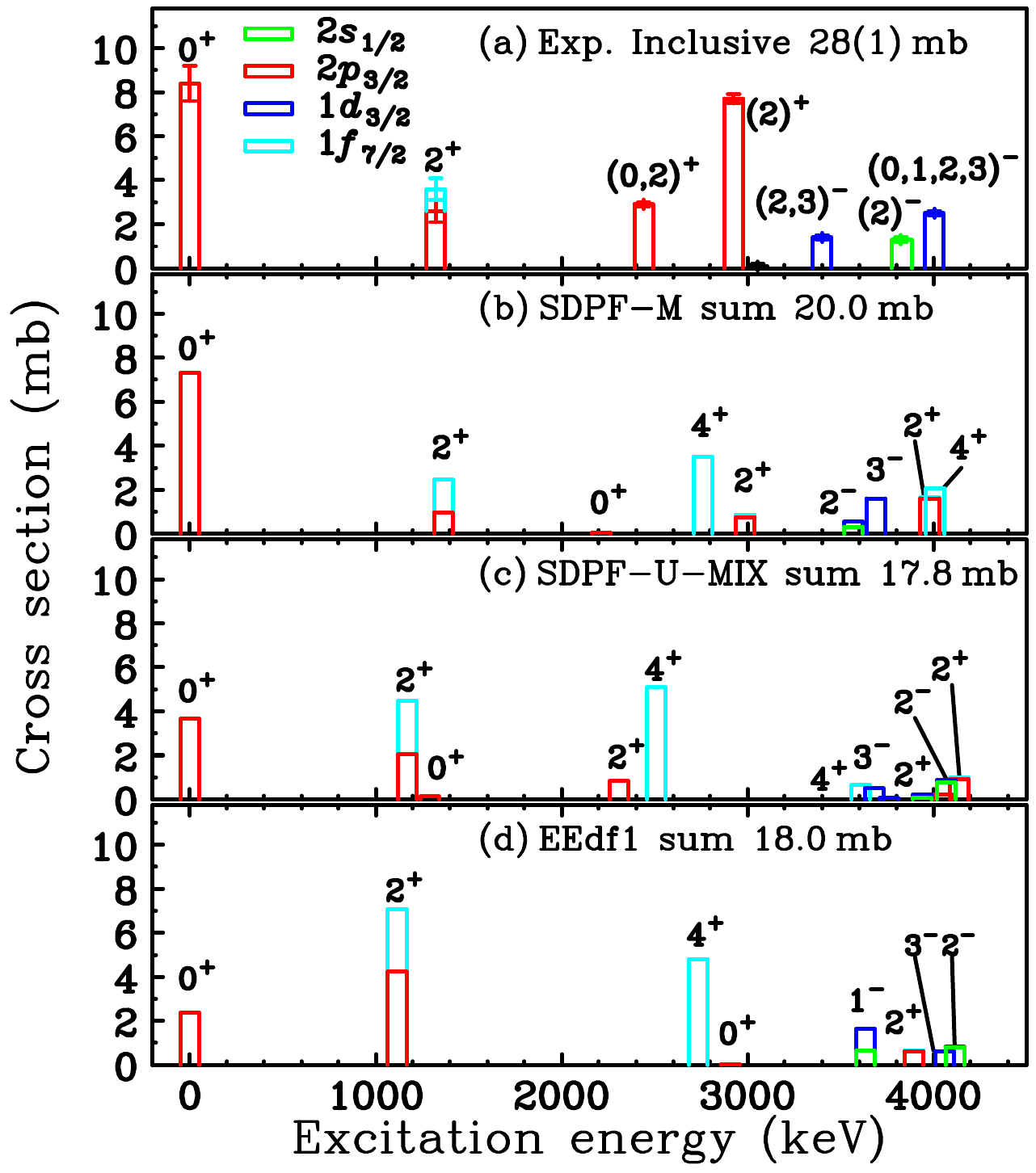}
\caption{\label{fig:xsec}
Comparison between (a) the measured partial cross sections and those calculated to the bound shell-model final states obtained using different interactions: 
(b) SDPF-M,  (c) SDPF-U-MIX, and (d) EEdf1.
The SDPF-M results are taken from Refs.~\cite{Kob29, eikonal_arxiv}.
Contributions of neutron removal from 2$s_{1/2}$,  2$p_{3/2}$, 1$d_{3/2}$, and 1$f_{7/2}$
orbitals are colored by green, red, blue, and cyan, respectively.
The single-particle cross sections are calculated using the DWIA framework.
The experimental ground-state cross section should be seen as an upper limit. 
The experimental inclusive cross section and the sum of calculated partial cross sections are also shown for each case.
}
\end{figure}

\begin{figure}
  \includegraphics[width=8.6cm]{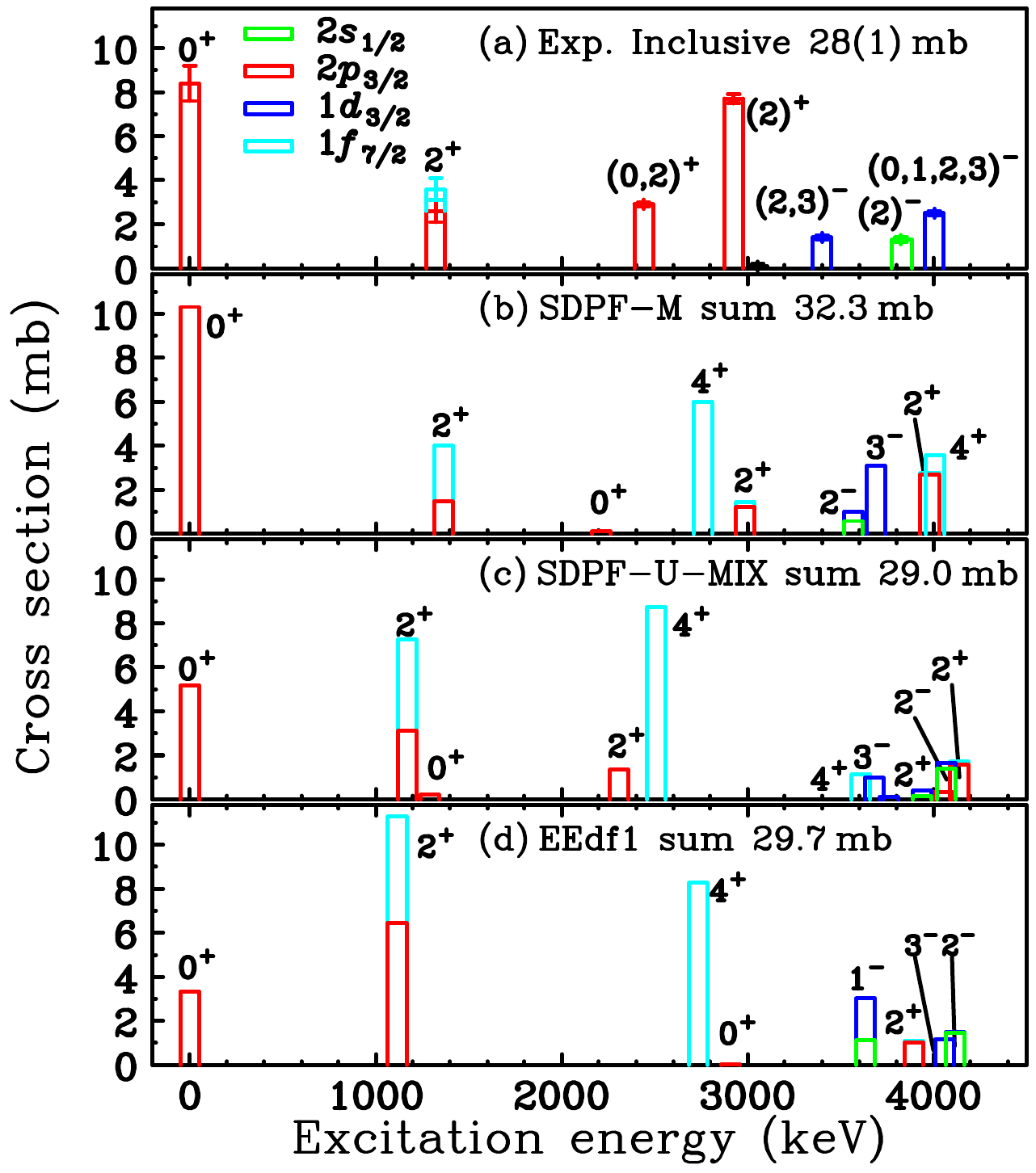}
\caption{\label{fig:xsec_eikonal}
Same as Fig.~\ref{fig:xsec} but with the removal cross sections calculated using the eikonal dynamical model.
}
\end{figure}

With the spin-parity assignments discussed, 
the experimental partial cross sections can be used to probe the wave function composition of the ground state of $^{29}$Ne. 
The experimental partial cross sections and those calculated using different shell-model spectroscopic factors are plotted in Fig.~\ref{fig:xsec} and Fig.~\ref{fig:xsec_eikonal}
using the DWIA framework~\cite{DWIA_1,DWIA_2} and the eikonal model~\cite{eikonal_arxiv}, respectively.
Evident from panel (a) of the two figures is the significant measured 2$p_{3/2}$-wave strength and the small observed 1$f_{7/2}$ contribution. 
The large 2$p_{3/2}$ contribution comes not only from the $0^+_{gs}$ and $2^+_1$ states, but also from the (0,2)$^+$ and the $2^+$ states above 2~MeV. 
Based on the measured partial cross sections, the summed experimental 2$p_{3/2}$ spectroscopic factors (C$^2$S$_{\rm exp}$ = $\sigma_{\rm exp}$/$\sigma_{\rm sp}$), 
using the DWIA and eikonal single-particle cross sections are 1.82(7) and 1.16(5), respectively, assuming the linear combination shown in Fig.~\ref{fig:momentum}(a)
(1.90(6) and 1.22(4) assuming a removal from 2$p_{3/2}$).
These values show a large summed $p$-wave strength compared to shell-model calculations as presented in Fig.~\ref{fig:xsec} and Fig.~\ref{fig:xsec_eikonal}. 
Such large occupations of the 2$p_{3/2}$ orbital shows its lowering, as compared to the normal 1$d_{3/2}$ orbital, 
indicating its intruder nature, thereby providing evidence of the breakdown of the $N=20$ and $N=28$ shell gaps~\cite{Liu_29Ne,SDPF-U-MIX,Doornenbal_38Mg}.

As shown above, the present work suggests weak possible $f$-wave intruder strengths to the bound states of $^{28}$Ne 
-- from the fit to the shape of the $2^+_1$ state momentum distribution.
Considering the $^{29}$Ne ground state is 3/2$^-$~\cite{Liu_29Ne, Kob29}, direct population of $4^+$ states in $^{28}$Ne is expected following a neutron removal from the intruder 1$f_{7/2}$ orbital,
if there were any significant $f$-wave occupancy.
The newly obtained data, however, do not show signs of $4^+$ states. 
Ref.~\cite{Fallon_28Ne} once proposed the 1732-keV transition as a candidate for the $4^+_1 \rightarrow 2^+_1$ decay. 
Even assuming the 3055-keV state is the $4^+_1$ state, the present data still suggests weak $f$-wave strength because of its small cross section of 0.1(1)~mb. 
This experimental finding in Ne is different from those in the Mg isotopes, 
where considerable $f$-wave strength to the bound states has been identified via one-neutron removal reactions~\cite{Terry_30Mg,Kitamura_30Mg,Bazin_33Mg,Kitamura_32Mg,Kitamura_32Mg_2}.
Such a difference in the $f$-wave occupancies between Ne and Mg might be related to the gap size between the 2$p_{3/2}$ and 1$f_{7/2}$ orbitals. 
For the Ne isotopes, shell-model calculations predict significant 1$f_{7/2}$-wave strength to bound final states in $^{27}$Ne~\cite{Brown_27Ne} and $^{29}$Ne~\cite{Liu_29Ne}. 
No such bound state strength of 1$f_{7/2}$ was reported experimentally~\cite{Terry_28Ne, Liu_29Ne}. 
In fact, the observed 1$f_{7/2}$-wave strength was found in unbound states at respective 1 and 1.5~MeV excitation above the 2$p_{3/2}$ states in $^{27}$Ne~\cite{Brown_27Ne} and $^{29}$Ne~\cite{Holl_29Ne},
leading to a gap between the 1$f_{7/2}$ and 2$p_{3/2}$ orbitals along the Ne isotopic chain approaching the island of inversion.
In contrast, the 1$f_{7/2}$ and 2$p_{3/2}$ orbitals are almost degenerate for Mg. 
Such a $pf$-orbital gap may suppress the occupation of the 1$f_{7/2}$ orbital in $^{29}$Ne, resulting in the weak $f$-wave strength to bound final states observed in $^{28}$Ne.

The excitation energies of the negative-parity states along the $N=18$ isotones can reflect the size of the $N=20$ gap approaching the island of inversion, 
as illustrated in Refs.~\cite{Kitamura_30Mg,Kitamura_32Mg}. 
Here, the negative-parity states in $^{28}$Ne are identified for the first time among the neutron-rich even-even Ne isotopes in the island of inversion.
Considering a 3p-4h configuration in $^{29}$Ne~\cite{Kob29,Revel_29Ne}, 
the large cross sections to the 3401- and 3825-keV states indicate the presence of the 3p-5h intruder configuration in $^{28}$Ne.
The lowest negative-parity state in $^{28}$Ne is found at 3.4~MeV, assuming a neutron removal from 1$d_{3/2}$.
This value is similar to that for $^{30}$Mg of 3.3~MeV, but is lower by about 1.8~MeV compared to $^{32}$Si, where a normal configuration dominates.
As shown in Fig.~\ref{fig:momentum}(d), the 3401-keV state is populated by removing the neutron from the 1$d_{3/2}$ orbital.
Thus, the decrease of the excitation energies of negative-parity states could be understood as 
a reduced gap between the 1$d_{3/2}$ and 2$p_{3/2}$ orbitals, thereby supporting the breakdown of the $N=20$ shell gap at $Z=10$.
Such a lowering of the negative-parity states is similar to that for the 1$^-$ state in $^{12}$Be along the $N=8$ isotones~\cite{Iwasaki_12Be}.

To seek insight from the theoretical description of $^{28}$Ne, 
large-scale shell-model calculations were performed using the SDPF-M~\cite{SDPF-M}, SDPF-U-MIX~\cite{SDPF-U-MIX}, and EEdf1~\cite{EEdf1} effective interactions. 
The SDPF-M interaction has been widely used in the studies for island-of-inversion nuclei. 
It includes a model space of the $sd$ shell and the 1$f_{7/2}$ and 2$p_{3/2}$ orbitals in the $pf$ shell, for both neutrons and protons.
SDPF-U-MIX is a more recently developed interaction with an extended model space including the full neutron $pf$ shell.
It has been successfully applied to describe the excited states for neutron-rich Mg and Si isotopes~\cite{SDPF-U-MIX}.
The recently developed EEdf1 interaction includes a model space consisting of the full $sd$ and $pf$ shells. 
Here, the two-body matrix elements are derived microscopically from the so-called extended Kuo-Krenciglowa (EKK) method.
The calculated level schemes using these three interactions are displayed in Fig.~\ref{fig:ne28_level} for comparison.
These three calculations show similar results for the $2^+_1$ and  $4^+_1$ excitation energies.
The calculated location of the $0^+_2$ state depends on the interaction. 
SDPF-U-MIX shows a $0^+_2$ state close to the $2^+_1$ state,
while the calculated $0^+_2$ excitation energies by SDPF-M and EEdf1 are located 1~MeV and 2~MeV above $2^+_1$, respectively.
All three calculations predict the negative-parity states to be located above 3~MeV.

To further characterize the intruder $pf$ spectroscopic strengths of the observed states,
the calculated partial cross sections are shown in Fig.~\ref{fig:xsec} and Fig.~\ref{fig:xsec_eikonal},
derived using the spectroscopic factors from these three shell-model interactions combined with the single-particle cross sections $\sigma_{sp}$ from the DWIA framework and the eikonal calculations.
The shell-model spectroscopic factors reflect the single-neutron overlaps between $^{28}$Ne and $^{29}$Ne, 
and thus depend not only on the final states of $^{28}$Ne but also on the initial (ground) state of $^{29}$Ne.
Both the SDPF-U-MIX and EEdf1 interactions predict a 3/2$^-$ as the $^{29}$Ne ground state 
with a large fraction of the 3p-4h configuration of 86\% and 83.8\%, respectively. 
For the SDPF-M calculation the $^{29}$Ne 3/2$^-$ state is not the ground state but lies at 73~keV~\cite{Liu_29Ne}. 
This state also shows a dominant 67\% 3p-4h configuration~\cite{Kob29, eikonal_arxiv}.
The summed calculated $p$ ($f$) spectroscopic factors are 
0.73 (0.86) for SDPF-M, 0.56 (0.96) for SDPF-U-MIX, and 0.81 (0.86) for EEdf1. 
In all three calculations, the $f$-wave spectroscopic strengths mainly come from the $4^+_1$ state.
Although SDPF-M and EEdf1 predict similar C$^2$S values for the $p$-wave, the main contributions are from the $0^+_{gs}$ and $2^+_1$, respectively.

As shown in Fig.~\ref{fig:xsec}(b)-(d) and Fig.~\ref{fig:xsec_eikonal}(b)-(d), all of the interactions qualitatively reproduce the distribution of $p$-wave strength, but underestimate their absolute values in general.
The SDPF-M result shows the most $p$-wave strength in the ground state
and the predicted ground-state cross section is in good agreement with the data. 
In the SDPF-U-MIX and EEdf1 calculations, the population of the ground state is hindered and the $p$-wave strength is shifted into the $2^+_1$ state.
In particular, the large $p$-wave strength found at 2923~keV is not present in any of the shell-model calculations.
All three calculations show a good agreement with the $2^+_1$ excitation energies.
We note that the $B(E2;0_{gs} \rightarrow 2^+_1)$ value of $^{28}$Ne is reproduced well by the EEdf1~\cite{EEdf1} calculation but is overestimated by the SDPF-M and SDPF-U-MIX results~\cite{Doornenbal_30Ne}.
For the $2^+_1$ partial cross section, not only $p$-wave but also sizable $f$-wave strength are suggested by the calculations, 
showing a marked difference from the experimental result, 
which shows a dominant $p$-wave strength as displayed in Fig.~\ref{fig:momentum}(a).
The EEdf1 calculation reproduces the $p$-wave strength to the $2^+_1$ state, but overestimates the cross section due to the existence of the theoretical $f$-wave strength.

Although the energy deviates, all three shell-model calculations predict very small C$^2$S values for the excited $0^+$ state, resulting in a small partial cross section.
Based on these calculated results, the 2441-keV state with its sizable cross section may not favor a $0^+$ assignment.
It is interesting to discuss the excitation energy of the $0^+_2$ state in $^{28}$Ne because excited $0^+$ states are important to reveal possible shape-coexistence.
Along the $N=18$ isotones, Ne is expected to have a greater intruder configuration than Mg, as Ne is expected to have a smaller $N=20$ shell gap~\cite{SDPF-M,SDPF-U-MIX}.
Thus, the $0^+_2$ state in $^{28}$Ne could be expected to appear lower in energy than that (at 1.8~MeV) for $^{30}$Mg~\cite{Schwerdtfeger_30Mg}. 
This scenario is supported by the SDPF-U-MIX result, which calculates the $0^+_2$ state to lie at 1.3~MeV. 
Similarly to $^{30,32}$Mg~\cite{Schwerdtfeger_30Mg,Wimmer_32Mg}, such a low excitation energy will lead to an isomeric $0^+_2$ state, presenting a shape-coexistence feature. 
A dedicated study of the excited $0^+$ state in $^{28}$Ne is therefore desired to reveal its nature -- which is a critical input for the overall theoretical description.

In contrast to the experimental findings, 
considerable 1$f_{7/2}$ contributions are presented in all the shell-model calculations, in particular for population of the $4^+_1$ state. 
In fact, along the Ne isotopic chain, the locations of the $f$-wave intruder states show disagreement between experimental results and shell-model calculations.
For instance, the $f$-wave intruder states were observed to be unbound in $^{27,29}$Ne~\cite{Brown_27Ne, Holl_29Ne}, while SDPF-M suggests they are bound.
In $^{29}$Ne, calculations with SDPF-U-MIX also show a bound $f$-wave intruder state at 610~keV, 
whereas the $f$-wave strength is observed at 1.5~MeV and unbound~\cite{Holl_29Ne,Liu_29Ne}.
Given the disagreements between the measured and shell-model expectations revealed by the present work, 
these $f$-wave intruder components present a challenge for the theoretical descriptions of the neutron-rich Ne isotopes. 
Moreover, a continued search for possible $f$-wave strength populating states in $^{28}$Ne is encouraged for future studies, 
for example, using $\gamma$-ray spectroscopy with high resolution and statistics to examine possible population of the $4^+_1$ state, 
as well as invariant-mass measurements to locate any unbound 4$^+$ states~\cite{Smith_28Ne} which will complement the experimental method used in the present work.

For the negative-parity states, all the shell-model calculations give a reasonable agreement of both the energies and their strength.
The reduction in the gap between the 1$d_{3/2}$ and 2$p_{3/2}$ orbitals suggested by the data, as discussed above, 
is also supported by the shell-model calculations~\cite{SDPF-M,SDPF-U-MIX,EEdf1}, which show the neutron $sd-pf$ gap narrows towards to $Z \le 10$.

In summary, intruder configurations in the ground state in $^{29}$Ne were investigated quantitatively for the first time 
via the one-neutron removal reaction on a hydrogen target.
The partial cross sections and the analysis of the momentum distributions reveal a strong $p$-wave strength to the low-lying states of $^{28}$Ne, 
but a weak, possible $f$-wave strength to bound $^{28}$Ne states.  
Such findings are different from the shell-model expectation that both considerable $p$- and $f$-wave strengths contribute to the bound states for the neutron-rich Ne isotopes, 
in a similar manner to the Mg isotopes.
The large value of summed $p$-wave spectroscopic factors presents clear evidence of the intruder nature of the 2$p_{3/2}$ orbital, 
as well as of the reduction of the $N=20$ and $N=28$ shell gaps.
The negative-parity states in $^{28}$Ne were identified for the first time, extending the systematics along the $N=18$ isotones, 
and once again supporting the breakdown of the $N=20$ shell gap in the neon isotopes. 
These general findings are common to both the DWIA framework and the eikonal dynamical model of the reaction.
Shell-model calculations using three different interactions SDPF-M, SDPF-U-MIX, and EEdf1 have been performed for the structure of $^{28}$Ne,  $^{29}$Ne, and their single-neutron overlaps.
However, none of these shell-model calculations reproduce the experimental observations of a large $p$-wave and a small $f$-wave strength,
thus presenting an ongoing challenge concerning the detailed theoretical description of the nuclei in the island of inversion.
Further work is also needed to assess the implications of the deduced spectroscopic strengths from these proton-target data, based on the
two presented theoretical reaction models, upon the analyses of the earlier published measurements of neutron-removal on C and Pb targets~\cite{Kob29}. 
The Coulomb dissociation data on the Pb target, being essentially independent of details of the strong-interaction removal-reaction mechanism, can provide an important benchmark of the observed differences between the deduced absolute intruder state strengths when using the two theoretical models.

\section*{Acknowledgments}
We express our gratitude to the RIKEN Nishina accelerator staff for providing the $^{48}$Ca beam and to the BigRIPS team for tuning the secondary beams. 
H.W acknowledges the support by MEXT KAKENHI Grant Nos. JP21H04465 and JP18H05404.
This work was supported partially by MEXT KAKENHI Grant No. JP16H02179. 
J.A.T acknowledges the support of the United Kingdom Science and Technology Facilities Council (STFC) Grant No. ST/V001108/1.
A.P. is supported by grants CEX2020-001007-S  funded by MCIN/AEI/10.13039/501100011033 and PID2021-127890NB-I00.
K.O acknowledges the support by MEXT KAKENHI Grant Nos. JP21H00125 and JP21H04975.
The LPC-Caen group wished to acknowledge support from the Franco-Japanese LIA-International Associated Laboratory for Nuclear Structure Problems as well as the French ANR-14-CE33-0022-02 EXPAND. 
This material is in part based upon work supported by the U.S. Department of Energy, Office of Science, Office of Nuclear Physics, under Contracts No. DE-AC02-06CH11357 (Argonne).
This work was in part supported by the Institute for Basic Science (IBS-R031-D1) in Republic of Korea. 
This work is partly supported from the National Research, Development and Innovation Fund of Hungary via Project No. K128947 and TKP2021-NKTA-42. 
This work was supported by the National Science Foundation, USA under Grants No. PHY-1102511.


\bibliography{ne28_V5}

\end{document}